\documentclass[aps,pra,twocolumn,showpacs,preprintnumbers,amsmath,amssymb,footinbib]{revtex4}
\usepackage{graphicx,epsfig}
\usepackage{braket}
\usepackage{bm}
\usepackage{dcolumn}
\usepackage{xcolor}
\usepackage[breaklinks=true,colorlinks,citecolor=blue,linkcolor=blue,urlcolor=blue]{hyperref}

\def\sp#1#2#3{Sci. Rep. {\bf #1}, #2 (#3)}

\def\prl#1#2#3{Phys. Rev. Lett. {\bf #1}, #2 (#3)}
\def\pra#1#2#3{Phys. Rev. A {\bf #1}, #2 (#3)}

\def\pre#1#2#3{Phys. Rev. E {\bf #1}, #2 (#3)}

\def\jpb#1#2#3{J. Phys. B: At. Mol. Opt. Phys. {\bf #1}, #2 (#3)}

\def\pla#1#2#3{Phys. Lett. A {\bf #1}, #2 (#3)}

\def\ajp#1#2#3{Am. J. Phys. {\bf #1}, #2 (#3)}

\def\noi{\noindent}
\def\bc{\begin{center}}
\def\ec{\end{center}}
\newcommand{\bea}{\begin{equation}}
\newcommand{\eea}{\end{equation}\noi}
\newcommand{\ber}{\begin{eqnarray}}
\newcommand{\eer}{\end{eqnarray}\noi}
\begin{document}
\title{Proof of quantum mechanical H-theorem beyond binary collisions in quantum gases}
\author{Bandita Das}
\author{Shyamal Biswas}\email{sbsp [at] uohyd.ac.in}
\affiliation{School of Physics, University of Hyderabad, C.R. Rao Road, Gachibowli, Hyderabad-500046, India
}

\date{\today}

\begin{abstract}
We have proved the quantum mechanical H-theorem for dilute Bose and Fermi gases by generalizing the quantum statistical Boltzmann equation for all possible many-body elastic collisions among the particles in the quantum gases within the Lippmann-Schwinger formalism. Previous study by Pauli did almost the same only for binary elastic collisions. We are considering all possible many-body elastic collisions for the current study. Our proof offers a better understanding to the foundation of the second law of thermodynamics for quantum gases. 
\end{abstract}


\maketitle
 

\section{Introduction}
Boltzmann's transport (kinetic) equation \cite{Boltzmann} and H-theorem \cite{Boltzmann} together form the statistical foundation of the second law of thermodynamics for a dilute gas of particles where binary collisions among the particles (atoms/molecules) play a leading role in reaching thermodynamic equilibrium of the system. Boltzmann's transport equation and H-theorem are based on the molecular chaos hypothesis (i.e. the velocities of colliding particles are uncorrelated and independent of their individual positions), the classical binary elastic collisions among the particles, and the classical diffusions of their probability density both in position space and momentum space \cite{Boltzmann}. Boltzmann's H-theorem says that, H-function ($H$) of a classical dilute gas, i.e. its negative entropy ($S$) in units of the Boltzmann constant ($H=-S/k_B$), never increases with time (i.e. $\frac{d H}{d t}=-\frac{1}{k_B}\frac{d S}{d t}\le0$ \cite{Boltzmann}) under the consideration of the molecular chaos hypothesis and the binary elastic collisions among the particles (atoms/molecules) in the dilute gas.

Later on, after the birth of quantum mechanics and quantum statistical mechanics, Pauli extended the H-theorem for quantum mechanical systems with the name-- quantum mechanical H-theorem or simply quantum H-theorem, and proved the same for binary elastic collisions in the dilute Bose and Fermi gases \cite{Pauli}. He introduced the quantum statistical Boltzmann equation (or simply, the quantum Boltzmann master equation \cite{Gardiner} or the quantum Boltzmann equation \cite{Vasilopoulos,Snoke}) and replaced the Boltzmann transport equation by the same specially to consider the binary elastic collisions in the quantum gases, and considered the quantum mechanical transitions within the single-particle states \cite{Pauli}. Random phase hypothesis which is the quantum analogue of the molecular chaos hypothesis is inbuilt in the quantum statistical Boltzmann equation \cite{Pauli,Pottier}. He also proved the quantum mechanical H-theorem even for a single-particle in the thermal radiation field by introducing the Pauli master equation, named after him \cite{Pauli}. Soon after this, von Neumann proved the quantum mechanical H-theorem by applying the Schrödinger equation (instead of the quantum statistical Boltzmann equation) with the consideration of macroscopic measurement procedure (as a counter to the molecular chaos hypothesis) for the particles in the quantum gases \cite{Neumann}. He also proved the quantum ergodic theorem \cite{Neumann}, in connection with the same, which was later attempted to be disproved by Farquhar and Landsberg \cite{Landsberg} resulting a controversy \cite{Goldstein} in the field.  After being inspired by the power of von Neumann's approach \cite{Neumann}, Han and Wu later realized the quantum mechanical H-theorem even without resorting to the macroscopic measurement procedure \cite{Han}. 

The quantum statistical Boltzmann equation is simple and plausible in realizing the approach to the equilibrium of a thermodynamically isolated (or even closed or open \footnote{The application needs fixing of the average energy for the closed system, and that of the average energy \& the average number of particles for the open system.}) system of a quantum gas. It directly connects quantum mechanics and thermodynamics. It goes beyond the first quantization and talks about the second quantization in the field theoretic language for the inter-particle collisions. Snoke \textit{et al} much later explicitly showed the connection of the quantum field theory and the quantum statistical Boltzmann equation in the language of many-body perturbation theory \cite{Snoke}. On the other hand, von Neumann's treatment on the quantum mechanical H-theorem, though powerful, does not directly enlight on how a thermodynamic system reaches equilibrium \cite{Neumann}.  Thus, the quantum statistical Boltzmann equation and Pauli's proof of the quantum mechanical H-theorem are widely accepted \cite{Tolman}. Moreover, one can derive Bose-Einstein or Fermi-Dirac statistics by imposing the condition of detailed balance on the quantum statistical Boltzmann equation in the thermodynamic equilibrium of a dilute quantum gas \cite{Tolman,Lawrence}.       

However, the molecular chaos hypothesis or its quantum analogue is not applicable to a microscopic system, such as quantum Maxwell demon \cite{Lioyd}, two-level system in the thermal radiation field, etc. One can expect aberrations of the second law of thermodynamics for such cases \cite{Lesovik,Biswas}. However, if the rates of stimulated emissions and absorptions are the same (+ve) and are time-independent, then no aberrations are found in the quantum mechanical H-theorem even for a two-level system (atom/molecule) in the thermal radiation field \cite{Feynman}. 

On the other hand, molecular chaos hypothesis or its quantum analogue is applicable to the macroscopic systems like a gas of particles in a macroscopic container of volume $V\rightarrow\infty$. Binary elastic collisions in a dilute quantum gas though take leading role in reaching the thermodynamic equilibrium, effects of three-body collisions are observed to be significant in making even a dilute Bose-Einstein condensate meta-stable in a 3-D macroscopic harmonic trap \cite{Cornell,Gammal,Kohler}. Thus, natural questions arise: how to generalize and prove the quantum mechanical H-theorem for three or more-body elastic collisions in dilute Bose and Fermi gases? Nobody answered these questions so far. This article is devoted to answer these questions within Pauli's framework not only for three-body elastic collisions but also for any number of many-body elastic collisions in dilute Bose and Fermi gases. 

Calculation in this article starts with the quantum statistical Boltzmann equation as introduced by Pauli for the binary elastic collisions in a dilute quantum (Bose or Fermi) gas. Then we revisit Pauli's proof of the quantum mechanical H-theorem for the binary elastic collisions in the dilute quantum gas. We also show an obvious extension of this proof with the various processes involved in the two-body collisions within the Lippmann-Schwinger formalism.  Then we generalize the master equation (i.e. the quantum statistical Boltzmann equation) for all possible many-body elastic collisions in the quantum gas within the Lippmann-Schwinger formalism. Then we prove the quantum mechanical H-theorem by applying the generalized master equation for dilute Bose and Fermi gases. Finally, we summarize our results in the conclusion.

\section{Proof of the quantum mechanical H-theorem}
Let us now revisit and generalize Pauli's proof of the quantum mechanical H-theorem for the Bose and Fermi gases \cite{Pauli} before coming to our proof beyond the binary collisions.

\subsection{Pauli's proof for binary elastic collisions}
Entropy of a free quantum gas away from its thermodynamic equilibrium at a temperature $T$ and chemical potential $\mu$ at time $t$ can be expressed in terms of the ensemble averaged occupation numbers ($\{\bar{n}_i(t)\}$) in its orthonormal and complete set of single-particle states ($\{\ket{\phi_i}\};~i=0,1,2,...$) as \cite{Pauli,Tolman} 
\begin{eqnarray}\label{eqn:1}
S(t)&=&-k_B\displaystyle\sum\limits_{i=0}^{\infty }\big[\bar{n}_i(t)\ln(\bar{n}_i(t))\nonumber\\&&\mp(1\pm\bar{n}_i(t))\ln(1\pm\bar{n}_i(t))\big]
\end{eqnarray}
where the upper sign stands for ideal Bose gas and the lower sign stands for ideal Fermi gas. According to the zeroth law of thermodynamics, the system starts evolving keeping its volume ($V\rightarrow\infty$) unaltered from $t\rightarrow-\infty$ and reaches the thermodynamic equilibrium adiabatically at a time $t\rightarrow\infty$ due to the attachment of the system to a heat and particle reservoir of temperature $T$ and chemical potential $\mu$.  The ensemble averaged occupation number $\bar{n}_i(t)$, in the thermodynamic limit,  takes the form of Bose-Einstein or Fermi-Dirac statistics in equilibrium as
\begin{eqnarray}\label{eqn:2}
\lim_{t\rightarrow\infty}\bar{n}_i(t)=\frac{1}{\text{e}^{(\epsilon_i-\mu)/k_BT}\mp1}
\end{eqnarray}
where $\epsilon_i$ is the energy of a particle in the single-particle state $\ket{\phi_i}$ and $\epsilon_0<\epsilon_1\le\epsilon_2\le....$. The form of equilibrium-entropy \cite{Pitaevskii,Kardar} is considered to be unaltered as in Eq.(\ref{eqn:1}) even when the ensemble averaged occupation number is time-dependent \cite{Pauli,Tolman,Feynman}. 

Effect of the reservoir on reaching the equilibrium of the system as well as on the evolution of the entropy of the system would be arbitrarily weak if we further fix its total averaged number of particles ($\sum_{i=0}^\infty\bar{n}_i(t)=\bar{N}=const.$) and the total averaged energy ($\sum_{i=0}^\infty\bar{n}_i(t)\epsilon_i=\bar{E}=const.$) on top of the fixed volume $V\rightarrow\infty$. The system, in this situation, essentially behaves like a thermodynamically isolated system, and evolution of its entropy ($S(t)$) due to that of $\bar{n}_i(t)$s would be arbitrarily slow in the thermodynamic limit. However, faster evolution of the entropy would be observed if we further consider even time-dependent constant perturbations in the system. 

Let us now consider weak (short ranged) inter-particle interactions, which are time-dependent constant perturbations in the system, be adiabatically turned on at time $t\rightarrow-\infty$ and turned off at $t\rightarrow\infty$. Thermodynamic equilibrium of the system would be reached faster, and the ensemble averaged occupation number, as in Eq.(\ref{eqn:2}), would be unaltered in this situation, as the interactions are switched off at a time faster than it reaches the equilibrium\footnote{See Ref.\cite{Feynman}, p. 1.}. Purpose of the H-theorem primarily is to explain time evolution the entropy of the system in the course of elastic collisions in it. Thus, H-function ($H(t)=-S(t)/k_B$) for the dilute quantum (Bose or Fermi) gas at a time $t$ can be expressed in terms of the ensemble averaged occupation numbers $\{\bar{n}_i(t)\}$ in its single-particle states $\{\ket{\phi_i(t)}'\}$ which are perturbed from $\{\ket{\phi_i}\}$ due to time-dependent constant perturbations in the course of elastic collisions, as \cite{Pauli,Tolman}     
\begin{eqnarray}\label{eqn:3}
H(t)=\displaystyle\sum\limits_{i=0}^{\infty}\big[\bar{n}_i(t)\ln(\bar{n}_i(t))\mp(1\pm\bar{n}_i(t))\ln(1\pm\bar{n}_i(t))\big].
\end{eqnarray}
Natural question arises: how the H-function for the thermodynamic system evolves with time? To know whether the H-function in Eq.(\ref{eqn:3}) is increasing or decreasing with time, we need to know only the first order time-derivative of the H-function
\begin{eqnarray}\label{eqn:4}
\frac{dH}{dt}=\displaystyle\sum\limits_{i=0}^{\infty }\frac{d\bar{n}_i}{dt}\big[\ln(\bar{n}_i)-\ln(1\pm\bar{n}_i)\big]
\end{eqnarray}
and the quantum statistical Boltzmann equation \cite{Pauli,Tolman,Lawrence}
\begin{eqnarray}\label{eqn:5}
\frac{d\bar{n}_i}{dt}&=&-\frac{2\pi}{\hbar}\displaystyle\sum\limits_j\frac{1}{2!}\displaystyle\sum\limits_{k,l}|\bra{i,j}\hat{V}_2\ket{k,l}'|^2\delta(\epsilon_i+\epsilon_j-\epsilon_k-\epsilon_l)\nonumber\\&&\times
\big[\bar{n}_i\bar{n}_j(1\pm\bar{n}_k)(1\pm\bar{n}_l)-\bar{n}_k\bar{n}_l(1\pm\bar{n}_i)(1\pm\bar{n}_j)\big].~~~~
\end{eqnarray}
Here, the summation is over all the two-body elastic collision events, in which, a particle in the unperturbed (initial\footnote{$t\rightarrow-\infty$}) state $\ket{\phi_i}$ scatters with another particle in the unperturbed state $\ket{\phi_j}$ into the scattering-out (perturbed) state $\ket{\phi_k}'$ or $\ket{\phi_l}'$ which is the time-independent part of $\ket{\phi_k(t)}'$ or $\ket{\phi_l(t)}'$ respectively. While the unperturbed states are plane waves, the scattering-out states are the linear combination of the plane waves and the radially outgoing waves (e.g. spherical waves) \cite{Englert}. Both the set of the unperturbed states and the scattering-out states are separately Dirac normalized and complete as we are not taking any bound states and scattering-in states (forward in time) under consideration \cite{Englert,Note}. While $\ket{i,j}=\frac{1}{\sqrt{2!}}[\ket{\phi_i}\ket{\phi_j}\pm\ket{\phi_j}\ket{\phi_i}]$ represents a two-body unperturbed state, $\ket{k,l}'=\frac{1}{\sqrt{2!}}[\ket{\phi_k}'\ket{\phi_l}'\pm\ket{\phi_k}'\ket{\phi_l}']$ represents a two-body scattering-out state \cite{Lawrence}. The term $\hat{V}_2$ represents the time-independent part of the two-body scattering matrix $\hat{V}_2(t)=\text{e}^{-\epsilon|t|/\hbar}\hat{V}_2$ where $\epsilon\rightarrow0^+$ \cite{Note2}. The factor $\frac{1}{2!}$ has appeared to avoid double counting of the norm-squared of the scattering matrix element ($|\bra{i,j}\hat{V}_2\ket{k,l}'|^2=|\bra{\phi_i}\bra{\phi_j}\hat{V}_2\ket{\phi_k}'\ket{\phi_l}'\pm\bra{\phi_i}\bra{\phi_j}\hat{V}_2\ket{\phi_l}'\ket{\phi_k}'|^2$) while the indices $k$ and $l$ are interchanged\footnote{The factor $\frac{1}{2!}$ in a number of references, e.g. in \cite{Tolman,Santos}, are absorbed in the norm-square of the matrix element.} \cite{Lawrence}. The norm-square of the scattering element is the same for the time reversed event for all $i,j,k,l$ by virtue of the microscopic time reversibility. The factors $\frac{2\pi}{\hbar}$, $|\bra{i,j}\hat{V}_2\ket{k,l}'|^2$ and $\delta(\epsilon_i+\epsilon_j-\epsilon_k-\epsilon_l)$ follow from Fermi's golden rule for the rate of transitions ($\Gamma_{i,j}^{k,l}=\frac{2\pi}{\hbar}|\bra{i,j}\hat{V}_2\ket{k,l}'|^2\delta(\epsilon_i+\epsilon_j-\epsilon_k-\epsilon_l)$ \cite{Fermi,Prugovecki,Note2}) of the two-body state $\{\ket{i,j}\}$ due to a single elastic collision. The term $\delta(\epsilon_i+\epsilon_j-\epsilon_k-\epsilon_l)$ corresponds to the elastic collision between the two particles. Fermi's golden rule needs the single-particle energy levels to be continuum and assumes the transition time to be negligibly small in comparison to the observation time scale. Thus Fermi's golden rule is equivalent to the random phase hypothesis and is compatible with the molecular chaos hypothesis. Finally, the term $\bar{n}_i\bar{n}_j(1\pm\bar{n}_k)(1\pm\bar{n}_l)$ represents the ensemble averaged number of the two-body collisions involving the unperturbed state $\ket{i,j}$ and the scattering-out state $\ket{k,l}'$ which reduces $\bar{n}_i$, and $\bar{n}_k\bar{n}_l(1\pm\bar{n}_i)(1\pm\bar{n}_j)$ represents the same for the time reversed event which increases $\bar{n}_i$ \cite{Santos}. Involvement of the population factor (i.e. last two terms) needs the assumption of no-correlations in the occupation numbers \cite{Snoke}. The upper sign (for bosons) and lower sign (for fermions) clearly say that the second quantization as well as a quantum field theory is involved in the quantum statistical Boltzmann equation though the quantum field theory was not well-built when Pauli phenomenologically arrived at the Eq.(\ref{eqn:5}) with the scattering-out state $\ket{k,l}'$ in it replaced by the unperturbed state $\ket{k,l}$ in 1928 \cite{Pauli}. 

Eq.(\ref{eqn:5}) is an obvious generalization of Pauli's first order perturbation result \cite{Pauli}, as because, the scattering-out state $\ket{k,l}'$ in it can be expressed within the Lippmann-Schwinger form as $\ket{k,l}'=\ket{k,l}+\frac{1}{(\epsilon_k+\epsilon_l)-\hat{H}_2^{(0)}+\text{i}\epsilon}\hat{V}_2\ket{k,l}'$ where $\hat{H}_{2}^{(0)}$ is the two-body Hamiltonian of the quantum gas in the ideal situation \cite{Prugovecki,Englert}.  Lippmann-Schwinger form of the scattering-out state in Eq.(\ref{eqn:5}) captures all the clustering features of the two-body collisions like those found in a virial expansion\footnote{Please see-- page no. 113 of Ref.\cite{Englert}.}. On top of this, we must mention that, the system evolves towards a diagonal state (many-body states with no phase coherences among the single-particle states) on the time scales comparable to the collision time provided the system starts with a purely diagonal state; otherwise the system may evolve away from the thermodynamic equilibrium \cite{Snoke}. Hence, we are assuming the system starts evolving from a pure diagonal state (which is highly probabilistic in general) and neglect contribution of the phase coherences in its entropy \cite{Snoke} for arriving at the quantum statistical Boltzmann equation.

Pauli proved the quantum mechanical H-theorem applying the quantum statistical Boltzmann equation (Eq.(\ref{eqn:5})) with $\ket{k,l}'\rightarrow\ket{k,l}$ by showing $\frac{d H}{dt}$ in Eq.(\ref{eqn:4}) always satisfies the inequality $\frac{d H}{dt}\le0$ \cite{Pauli}. However, Pauli considered only binary elastic collisions in the master equation (Eq.(\ref{eqn:5})). Thus, his proof is valid only for binary elastic collisions in the dilute Bose and Fermi gases. Though binary collisions play a leading role in reaching the equilibrium state, effect of three-body and more-body elastic collisions are not always negligible even in a dilute gas \cite{Cornell,Gammal,Kohler}. Tolman attempted to discuss the role of the three-body elastic collisions in the Boltzmann's (classical) H-theorem in Ref. \cite{Tolman2}. However, role of the three-body and more-body elastic collisions has not been surprisingly discussed so far for the validity of the quantum mechanical H-theorem. We are offering a proof of the quantum mechanical H-theorem in the following for all possible many-body elastic collisions in the dilute Bose and Fermi gases.     

\subsection{Proof for all possible many-body elastic collisions}
We generalize the quantum statistical Boltzmann equation (Eq.(\ref{eqn:5})) for  all possible many-body elastic collisions in a dilute quantum (Bose or Fermi) gas, with all the assumptions mentioned in the previous subsection, as
\begin{eqnarray}\label{eqn:6}
\frac{d\bar{n}_{i_1}}{dt}=\frac{d\bar{n}_{i_1}^{(2)}}{dt}+\frac{d\bar{n}_{i_1}^{(3)}}{dt}+...+\frac{d\bar{n}_{i_1}^{(k)}}{dt}+... ,
\end{eqnarray}
where the left hand side represents the rate of change in the ensemble averaged occupation number in the single-particle scattering-out state $\ket{\phi_{i_1}(t)}'$, the first term $\frac{d\bar{n}_{i_1}^{(2)}}{dt}$ in the right hand side corresponds to the known quantum statistical Boltzmann equation in Eq.(\ref{eqn:5}) with the dummy variable $i$ replaced by another dummy variable $i_1$, $k$ takes the values $2,3,4,...$, and $\frac{d\bar{n}_{i_1}^{(k)}}{dt}$ represents the contribution of $k$-body elastic collisions to the rate of change in the ensemble averaged occupation number in the scattering-out state $\ket{\phi_{i_1}(t)}'$ as
\begin{eqnarray}\label{eqn:7}
\frac{d\bar{n}_{i_1}^{(k)}}{dt}&=&-\frac{1}{(k-1)!k!}\sum_{i_2,...,i_k;j_1,j_2,...,j_k}\Gamma_{i_1,i_2,...,i_k}^{j_1,j_2,...,j_k}\nonumber\\&&\times\bigg[\mathcal{N}_{\bar{n}_{i_1},\bar{n}_{i_2},...,\bar{n}_{i_k}}^{\bar{n}_{j_1},\bar{n}_{j_2},...,\bar{n}_{j_k}}-\mathcal{N}_{\bar{n}_{j_1},\bar{n}_{j_2},...,\bar{n}_{j_k}}^{\bar{n}_{i_1},\bar{n}_{i_2},...,\bar{n}_{i_k}}\bigg]
\end{eqnarray}
where each of all the indices takes values from $0$ to $\infty$ with unit interval, $\frac{1}{(k-1)!}$ has come due to the symmetry (for bosons) or antisymmetry (for fermions) of the unperturbed state to avoid counting $(k-1)!$ times while summing over the indices $i_2,i_3,...,i_k$, the factor $\frac{1}{k!}$ has come due to the symmetry or antisymmetry of the scattering-out state to avoid counting $k!$ times while summing over the indices $j_1,j_2,...,j_k$,
\begin{eqnarray}\label{eqn:8}
\Gamma_{i_1,i_2,...,i_k}^{j_1,j_2,...,j_k}&=&\frac{2\pi}{\hbar}|\bra{i_1,i_2,...,i_k}\hat{V}_k\ket{j_1,j_2,...,j_k}'|^2\times\nonumber\\&&\delta(\epsilon_{i_1}+\epsilon_{i_2}+...+\epsilon_{i_k}-\epsilon_{j_1}-\epsilon_{j_2}-...-\epsilon_{j_k})~~~~
\end{eqnarray}
represents rate of transitions per $k$-body elastic collisions from the $k$-body normalized unperturbed state $\ket{i_1,i_2,...,i_k}$ to the $k$-body scattering-out state $\ket{j_1,j_2,...,j_k}'$, $\epsilon_{i_q}$ ($\epsilon_{j_q}$) is the energy in the time-independent single-particle state $\ket{\phi_{i_q}}$ ($\ket{\phi_{j_q}'}$) before (after) collision for $q=1,2,...,k$ belonging to an orthonormalized and complete set of the single-particle states $\{\ket{\phi_{i_q}}\}$ ($\{\ket{\phi_{j_q}}'\}$), $\hat{V}_k$ represents time-independent part of the $k$-body scattering matrix ($\hat{V}_k(t)=\text{e}^{-\epsilon|t|/\hbar}\hat{V}_k$), and finally
\begin{eqnarray}\label{eqn:9}
\mathcal{N}_{\bar{n}_{i_1},\bar{n}_{i_2},...,\bar{n}_{i_k}}^{\bar{n}_{j_1},\bar{n}_{j_2},...,\bar{n}_{j_k}}=\bar{n}_{i_1}\bar{n}_{i_2}...\bar{n}_{i_k}(1\pm\bar{n}_{j_1})(1\pm\bar{n}_{j_2})...(1\pm\bar{n}_{j_k})~
\end{eqnarray}
represents the ensemble averaged number of $k$-body elastic collisions involving the unperturbed state $\ket{i_1,i_2,...,i_k}$ and the scattering-out state $\ket{j_1,j_2,...,j_k}'$ which decrease the occupation in the single-particle state $\ket{\phi_{i_1}(t)}$ and $\mathcal{N}_{\bar{n}_{j_1},\bar{n}_{j_2},...,\bar{n}_{j_k}}^{\bar{n}_{i_1},\bar{n}_{i_2},...,\bar{n}_{i_k}}$ accounts for the time reversed events which increase the occupation in the single-particle state $\ket{\phi_{i_1}(t)}$. 

Now, considering microscopic time reversal symmetry in the elastic collisions (i.e. $|\bra{i_1,i_2,...,i_k}\hat{V}_k\ket{j_1,j_2,...,j_k}'|^2=|\bra{j_1,j_2,...,j_k}\hat{V}_k\ket{i_1,i_2,...,i_k}'|^2$), we recast Eq.(\ref{eqn:4}) with the generalized quantum statistical Boltzmann equation in Eq.(\ref{eqn:6}) as
\begin{eqnarray}\label{eqn:10}
\frac{dH}{dt}=\frac{dH^{(2)}}{dt}+\frac{dH^{(3)}}{dt}+...+\frac{dH^{(k)}}{dt}+...
\end{eqnarray}
where $\frac{dH^{(k)}}{dt}$ represents contribution of the $k$-body elastic collisions to the rate of change of the H-function, and is given by
\begin{eqnarray}\label{eqn:11}
\frac{dH^{(k)}}{dt}&=&-\sum_{i_1,i_2,...,i_k;j_1,j_2,...,j_k}\frac{\Gamma_{i_1,i_2,...,i_k}^{j_1,j_2,...,j_k}}{(k-1)!k!2^k}\bigg[\mathcal{N}_{\bar{n}_{i_1},\bar{n}_{i_2},...,\bar{n}_{i_k}}^{\bar{n}_{j_1},\bar{n}_{j_2},...,\bar{n}_{j_k}}\nonumber\\&&-\mathcal{N}_{\bar{n}_{j_1},\bar{n}_{j_2},...,\bar{n}_{j_k}}^{\bar{n}_{i_1},\bar{n}_{i_2},...,\bar{n}_{i_k}}\bigg]\bigg[\ln\bigg(\mathcal{N}_{\bar{n}_{i_1},\bar{n}_{i_2},...,\bar{n}_{i_k}}^{\bar{n}_{j_1},\bar{n}_{j_2},...,\bar{n}_{j_k}}\bigg)\nonumber\\&&-\ln\bigg(\mathcal{N}_{\bar{n}_{j_1},\bar{n}_{j_2},...,\bar{n}_{j_k}}^{\bar{n}_{i_1},\bar{n}_{i_2},...,\bar{n}_{i_k}}\bigg)\bigg].
\end{eqnarray}
Here, the factor $\frac{1}{2^k}$ has come by virtue of the microscopic time reversal symmetry to avoid $2^k$ counting while interchanging $i_1$ with each of $j_1,j_2,...,j_{k-1}$ and $j_k$. It is interesting to note that  $\Gamma_{i_1,i_2,...,i_k}^{j_1,j_2,...,j_k}$ in Eq.(\ref{eqn:11}) is nonnegative by definition, and if $\mathcal{N}_{\bar{n}_{i_1},\bar{n}_{i_2},...,\bar{n}_{i_k}}^{\bar{n}_{j_1},\bar{n}_{j_2},...,\bar{n}_{j_k}}\ge\mathcal{N}_{\bar{n}_{j_1},\bar{n}_{j_2},...,\bar{n}_{j_k}}^{\bar{n}_{i_1},\bar{n}_{i_2},...,\bar{n}_{i_k}}$ then $\ln\bigg(\mathcal{N}_{\bar{n}_{i_1},\bar{n}_{i_2},...,\bar{n}_{i_k}}^{\bar{n}_{j_1},\bar{n}_{j_2},...,\bar{n}_{j_k}}\bigg)-\ln\bigg(\mathcal{N}_{\bar{n}_{j_1},\bar{n}_{j_2},...,\bar{n}_{j_k}}^{\bar{n}_{i_1},\bar{n}_{i_2},...,\bar{n}_{i_k}}\bigg)\ge0$, and if $\mathcal{N}_{\bar{n}_{i_1},\bar{n}_{i_2},...,\bar{n}_{i_k}}^{\bar{n}_{j_1},\bar{n}_{j_2},...,\bar{n}_{j_k}}\le\mathcal{N}_{\bar{n}_{j_1},\bar{n}_{j_2},...,\bar{n}_{j_k}}^{\bar{n}_{i_1},\bar{n}_{i_2},...,\bar{n}_{i_k}}$ then $\ln\bigg(\mathcal{N}_{\bar{n}_{i_1},\bar{n}_{i_2},...,\bar{n}_{i_k}}^{\bar{n}_{j_1},\bar{n}_{j_2},...,\bar{n}_{j_k}}\bigg)-\ln\bigg(\mathcal{N}_{\bar{n}_{j_1},\bar{n}_{j_2},...,\bar{n}_{j_k}}^{\bar{n}_{i_1},\bar{n}_{i_2},...,\bar{n}_{i_k}}\bigg)\le0$. This implies 
\begin{eqnarray}\label{eqn:12}
\frac{dH^{(k)}}{dt}\le0
\end{eqnarray}
all times for all values of $k=2,3,4,...$, and thus we set the inequality to the left hand side of Eq.(\ref{eqn:10}) as
\begin{eqnarray}\label{eqn:13}
\frac{dH}{dt}\le0.
\end{eqnarray}
Hence, the quantum mechanical H-theorem is proved for all possible many-body elastic collisions in dilute Bose (upper sign) and Fermi (lower sign) gases.    

\section{Conclusions}
We have proved the quantum mechanical H-theorem for dilute Bose and Fermi gases by generalizing the quantum statistical Boltzmann equation for all possible many-body elastic collisions among the particles in the quantum gases within the random phase hypothesis (which is equivalent to the molecular chaos hypothesis) and the assumptions of no occupation number correlations,  no many-body phase coherence, etc.  We have clearly shown the individual role of all possible many-body elastic collisions in the quantum mechanical H-theorem. We exactly get back Pauli's proof \cite{Pauli} from our general proof in this regard by setting $\hat{V}_3=\hat{V}_4=\hat{V}_5=...=0$ and neglecting the higher orders of $\ket{j_1,j_2}'=\ket{j_1,j_2}+\frac{1}{(\epsilon_{j_1}+\epsilon_{j_2})-\hat{H}_2^{(0)}+\text{i}0^+}\hat{V}_2\ket{j_1,j_2}'$ in the Eq.(\ref{eqn:13}). We also get back classical (Boltzmann's) H-theorem at a very high temperature ($T\rightarrow\infty$). 

Quantum mechanical H-theorem is often expressed in terms of degeneracies ($g_i$) of the single-particle energy levels ($\epsilon_i$) with summation over non-degenerate levels both in the quantum statistical Boltzmann equation and the H-function \cite{Pauli,Tolman}. Here, we have considered degeneracy over the ground state by setting $\epsilon_0<\epsilon_1\le\epsilon_2\le....$ and by considering summation over all the single-particle states. If one wants to express the thermodynamic quantities in terms of degeneracies and summation over the single-particle every levels, then $1\pm\bar{n}_i$ is to be replaced by $g_i\pm\bar{n}_i$ for all the single-particle states everywhere in this article.

Whether the H-theorem would be valid for three-body or more-body elastic collisions was clear neither from Pauli's proof of the quantum mechanical H-theorem \cite{Pauli} nor from its subsequent discussions in the literature \cite{Tolman,Santos,Gemmer}. We have explicitly shown it in this article. Our proof offers a better understanding to the foundation of the second law of thermodynamics for quantum gases.

We have shown the quantum mechanical H-theorem to be valid for many-body elastic collisions with instantaneous repulsive (or attractive) interactions among the particles of a quantum (Bose or Fermi) gas in the thermodynamic limit. Thus, it is necessary to take an energy conserving system of a dilute gas, i.e. a thermodynamically isolated system, for the validity of the H-theorem. If we take a thermodynamically closed system, then we have to ensure its average energy be kept constant; and if we take a thermodynamically open system, then we have to ensure its average energy and average number of particles be kept constant for the validity of the H-theorem. Thus, entropy ($S=-k_BH$) is a non-decreasing function under many-body elastic collisions in such a system. We have considered the system to be evolved from a diagonal state, for proving the quantum mechanical H-theorem, so that the many-body phase coherences in the system would be neglected all the time \cite{Snoke}. 

The quantum mechanical H-theorem is not valid for microscopic systems where the molecular chaos hypothesis (or its equivalent hypothesis) is not applicable \cite{Lesovik,Biswas}. It is also not valid for dissipative systems for inelastic collisions in them \cite{Sackett}. It is also not valid for the interacting Bose-Einstein condensates as the assumption of `no occupation number correlations' is not applicable there \cite{Snoke}. 

Equality in Eq.(\ref{eqn:13}) holds in the thermodynamic equilibrium, which in turn sets detailed balance and results Bose-Einstein or Fermi-Dirac statistics as in Eq.(\ref{eqn:2}). From Eq.(\ref{eqn:13}) one can infer that, total entropy of the system in equilibrium is sum of the entropies due to the two-body elastic collisions, three-body elastic collisions and so on over the entropy due to no collisions (which would be the constant of integration). 

Our proof of the quantum mechanical H-theorem is limited to the dilute quantum gases. Bound states may be formed in the process of collisions with attractive interactions. This would cause reduction of entropy. We have not considered such processes, though they are very less probabilistic in the dilute quantum gases, for proving the H-theorem. Formation of bound states would be more probabilistic for strongly interacting cases with attractive interactions. H-theorem is not applicable in this situation. Tuning of the H-function with the coupling constant for the interactions is kept as an open problem.

\acknowledgments
S. Biswas acknowledges financial support of the DST, Govt. of India under the INSPIRE Faculty Award Scheme [No. IFA-13 PH-70]. Useful discussions with Dr. Saugata Bhattacharyya (Vidyasagar College, Kolkata) and Prof. J. K. Bhattacharjee (IACS, Kolkata) are gratefully acknowledged.

\end{document}